\input psbox.tex
\documentstyle[12pt]{article}
\normalsize

\def\bar#1{\overline{#1}}

\def\Hat#1{\rlap{\kern.10em$\widehat{\phantom G}$}#1}
\def\HAt#1{\rlap{\kern.05em$\widehat{\phantom G}$}#1}

\def\cap#1{\rlap{\kern.1em$\widehat{\phantom{G\vrule height.8em}}$}#1{}}
\def\Cap#1{\rlap{\kern.05em$\widehat{\phantom{G\vrule height.8em}}$}#1{}}

\let\oldtheequation=\theequation
\def\doteqs#1{\setcounter{equation}{0}
            \def\theequation{{#1}.\oldtheequation}}
\newcounter{sxn}
\def\sx#1{\addtocounter{sxn}{1} \bigskip\medskip \goodbreak
\noindent{\large\bf\centerline{\thesxn.~~#1}} \nobreak \medskip}
\def\sxn#1{\sx{#1} \doteqs{\thesxn}}

\newcounter{axn}

\def\br{\begin{thebibliography}{abc}}
\def\er{\end{thebibliography}}

\date{}

\begin{document}
\bibliographystyle{unsrt}
\footskip 1.0cm
\thispagestyle{empty}
\setcounter{page}{0}
\begin{flushright}
IMSc-93/32\\
SU-4240-538\\
September, 1993 \\
\end{flushright}
\vspace{10mm}

\centerline {\LARGE MAXWELL-CHERN-SIMONS}
\vspace{5mm}
\centerline {\LARGE ELECTRODYNAMICS ON A DISK}
\vspace*{15mm}
\centerline {\large A.P. Balachandran$^{(1)}$, L. Chandar$^{(1)}$,
E. Ercolessi$^{(1)}$}
\centerline {\large T.R. Govindarajan$^{(2)}$ and R. Shankar$^{(2)}$}

\vspace*{5mm}
\centerline {$^{(1)}$\it Department of Physics, Syracuse University,}
\centerline {\it Syracuse, NY 13244-1130, U.S.A.}
\centerline {$^{(2)}$\it Institute of Mathematical Sciences,}
\centerline {\it Madras, T.N. 600 113, India}
\vspace*{15mm}
\normalsize
\centerline {\bf Abstract}
\vspace*{5mm}

The Maxwell-Chern-Simons (MCS) Lagrangian is the Maxwell Lagrangian
augmented by the Chern-Simons (CS) term. In this paper, we study the
MCS and Maxwell Lagrangians on a disk $D$. They are of interest
for the quantum Hall effect, and also when the disk and its exterior
are composed of different media. We show that quantization is not
unique, but depends on a nonnegative parameter $\lambda$. $1/\lambda$
is the penetration depth of the fields into the medium
in the exterior of
$D$. For $\lambda=0$, there are edge observables and edge states
localized at the boundary $\partial D$ for the MCS system. They
describe the affine Lie group $\tilde{L}U(1)$. Their excitations
carry zero energy signifying an infinite degeneracy of all states of
the theory. There is also an additional infinity of single particle
excitations of exactly the same energy proportional to $|k|$, $k$
being the strength of the CS term. The MCS theory for $\lambda=0$ has
the huge symmetry group $\tilde{L}U(1) \times U(\infty)$. In the Maxwell
theory, the last mentioned excitations are absent while the edge
observables, which exist for $\lambda=0$, commute. Also these
excitations are described by states which are not localized at $\partial
D$ and are characterized by a continuous and infinitely
degenerate spectrum. All these degeneracies are lifted and edge
observables and their states cease to exist for $\lambda>0$.

The novel excitations discovered in this paper should be accessible
to observations. We will discuss issues related to observations, as
also the generalization of the present considerations to vortices,
domain walls and monopoles, in a paper under preparation.

\newpage

\baselineskip=24pt
\setcounter{page}{1}
\newcommand{\be}{\begin{equation}}
\newcommand{\ee}{\end{equation}}

\sxn{Introduction}

In spacetime of 2+1 dimensions, the conventional Maxwell Lagrangian
for free electrodynamics can be augmented by the Chern-Simons term without
spoiling gauge invariance.  The result describes a modification of
electrodynamics \cite{1} which may be called Maxwell Chern-Simons (MCS)
electrodynamics.
In the past, it has been studied when the underlying
spatial manifold $M^{2}$ is compact or noncompact and many interesting
physical results have been obtained.  It has in particular  been shown
to describe a massive free field when $M^{2}$ is ${\bf R}^{2}$, whereas the
field is massless in the absence of the Chern-Simons term \cite{1}.
Speculations have also been advanced that analogues of this mass generation
mechanism in higher dimensions may be viable alternatives to the well known
approach to vector meson mass employing the Higgs field \cite{2}.

In this paper, we examine MCS and Maxwell electrodynamics when $M^{2}$ is
a disc $D$.  There are good physical reasons for undertaking this
task.  It is widely felt that the Chern-Simons term and its
variants are of fundamental importance for the quantum Hall effect.
In particular, when the Hall sample is confined to $D$, one knows from
microscopic theory (\cite{3}, for a review and references, cf.\ ref.\
\cite{4}) that
there are edge states localized on the boundary \mbox{$\partial D$} of
$D$, and the Chern-Simons term and its modifications, including
MCS electrodynamics, seem well adapted to describe these states (cf.\ ref.\
\cite{5}).
Further, we will establish several specific results below associated with
our finite geometry, and they will depend on a nonnegative parameter
$\lambda$ with the dimension of mass.  Its existence and consequences do
not seem to have been appreciated in previous work.  It has a good
interpretation as well: in a second paper \cite{6}, we will argue
that it will naturally arise if the disk is surrounded by a superconductor,
$1/\lambda$ being the penetration depth of the latter.  All
this seems
new and also provides justification for studying the MCS and Maxwell
electrodynamics on a disk, the latter being just a special case of the former
for zero Chern-Simons term.  A final reason for
this work can be the following.  Edge states are described by conformal
field theories and affine Lie groups \cite{5,7,8} and form the bridge between
three-dimensional gauge theories and two-dimensional conformal field theories.
It is therefore important to try to understand them as much as possible.

Previous papers on edge states, including our own \cite{5}, relied
on semiclassical or path integral approaches for their derivation.
In contrast, in this paper, we will pay attention to domain problems of
operators.  While preoccupation with domains of operators for many physical
systems adds to little else but rigour, that appears not to be the case when
dealing with manifolds having boundaries.  Thus, in the cases we consider here,
there are
many inequivalent quantizations depending on our choice of domain (or
``boundary conditions'') for a certain second order operator.  In this
paper, we study only boundary conditions compatible with locality and
angular momentum conservation.  They
depend on the parameter $\lambda$ alluded to above, each $\lambda$ giving
a distinct quantum theory.  All previous work we are familiar with effectively
assumes that $\lambda = 0$.

Let \mbox{$1/e^{2}$} and $k$ be the constants characterizing the strengths of
the Maxwell and Chern-Simons terms.  The electrodynamics we consider  on $D$
then depends on the three constants \mbox{$1/e^{2}$}, $k$ and $\lambda$.  In
this paper, we will treat the $k\neq 0$ and $k=0$ cases separately and find
the following results:
\\
\underline{For $k\neq 0$}
\begin{enumerate}
\item Edge observables and edge states exist only when $\lambda = 0$. All
observables and states are spread out over the disk when $\lambda$ is different
from zero.
\item The centrally extended loop group \mbox{$\tilde{L}U(1)$} \cite{7,8}
found in previous work for $\lambda = 0$ also does not seem to exist if
$\lambda
\neq 0$.  Thus, although the generators of this group can be
formally written down, its action apparently fails to preserve the domain of
the
Hamiltonian and is anomalous in the same way that axial transformations in
QCD are anomalous \cite{9}.
\item The second quantized theory can be explicitly solved for $\lambda = 0$
and has the following states from which the Fock states
can be built up.  There are first the massive photon modes.  There are also
the edge states localized at \mbox{$\partial D$} all of which have zero
energy.  There is \underline{in addition} an infinite family of
exactly degenerate states which are \underline{not} localized at the edge and
with energy proportional to $|k|$.  They are thus split from vacuum for
nonzero $k$.  They will be called
harmonic states for reasons which will become clear in Section 4, and will
be denoted by \mbox{$H_{n}(0)$}.  Here $n$ is a nonzero integer and 0 is the
value of $\lambda$.
\item When $\lambda$ is deformed away from zero, the edge states cease to
exist with localization at the edge.  Also the
deformations \mbox{$H_{n}(\lambda )$} of the harmonic states which were
degenerate earlier for $\lambda =0$ are now split in energy
from each other.
\end{enumerate}
\underline{For $k=0$ or for Maxwell's theory}
\begin{description}
  \item[ a)] Just as for $k\neq 0$, edge observables exist only for $\lambda
=0$, all observables being spread out over the disk when $\lambda$ deviates
from zero.
  \item[ b)] The edge observables of $\lambda =0$ are an infinite number of
\underline{commuting} constants of motion so that they do not describe a
conformal field theory or the group $\tilde{L}U(1)$.
  \item[ c)] The variables conjugate to the edge observables and the states
associated with both are \underline{not} localized at the edge even for
$\lambda =0$.  This is unlike the situation for $k\neq 0$.
  \item[ d)] The second quantized theory can be explicitly solved for any value
of $\lambda$.  For $\lambda =0$, the degenerate states with mass proportional
to $|k|$ mentioned previously become the same as the states associated with the
edge observables as $k\rightarrow 0$.  But they do not have zero energy, but
rather a continuous spectrum of energies from 0 to $\infty$, each point of this
spectrum
being infinitely degenerate.  There are in addition the massless photon modes.
The space of states of quantum field theory are obtained from suitable tensor
products of these states.
  \item[ e)] This degeneracy and continuous spectrum cease to exist when
$\lambda$ becomes nonzero.  There are no edge observables either in that case
as mentioned previously.
\end{description}

It should be clear from these remarks that the $k\rightarrow 0$ limit of the
MCS theory on $D$ is not smooth.

The paper is organized as follows.

In Section 2, we establish the boundary conditions necessary for quantization
and show how they depend on $\lambda$.

We specialize to $\lambda = 0$ in Sections 3 and 4.  We also assume that $k\neq
0$ in Sections 4 and 5.

  In Section 3, we present
transparent classical arguments which clearly suggest the presence of
degenerate harmonic states with energy proportional to $|k|$ in quantum
theory when $k$ differs from zero.
It is also pointed out that the classical modes associated with these states
have energy when $k=0$.

The boundary conditions of Section 2 are for a second order differential
operator.  Section 4 develops a Hodge theory \cite{10} to explicitly solve
for the eigenfunctions and spectrum of this operator, thereby explicitly second
quantizing our Lagrangian as well.  All the results claimed under 3 above are
established.

Section 5 considers nonzero $\lambda$.  The edge states of $\lambda
= 0$ do not exist when $\lambda$ deviates from zero.  Further
the deformed harmonic states are discovered to be no longer
mutually degenerate.  Their actual energies can
perhaps be calculated numerically, but we have not attempted that task here,
not having succeeded in developing a scheme for diagonalizing the Hamiltonian.
In this Section we also show that the affine Lie group, present as a group of
symmetries for $\lambda =0$, ceases to exist for $\lambda \neq 0$.

Section 6 deals with Maxwell's theory and establishes its features described
previously.

\sxn{Boundary Conditions for Quantization and the Parameter $\lambda$}

Throughout this paper, we will use the following conventions:
\begin{enumerate}
\item Greek and Latin indices take values 0,1,2 and 1,2 respectively.
\item The three-dimensional metric $\eta$ is specified by its nonvanishing
entries \\$\eta _{00}=-1,\;\; \eta _{11}=\eta _{22}=+1$ while the
three-dimensional Levi-Civita symbol is \\$\epsilon _{\mu \nu \lambda}$
with $\epsilon _{012}=+1$.
\item The spatial metric is given by the Kronecker $\delta$ symbol while the
two-dimensional Levi-Civita symbol is $\epsilon _{ij}$ with $\epsilon
_{12}=+1$.
\end{enumerate}

We will also assume that the disk has a circular boundary and radius $R$.

The Lagrangian we plan to study here is
\begin{eqnarray}
L & = & \int d^{2}x \:{\cal L}, \nonumber    \\
{\cal L} & = & -\frac{1}{4e^{2}} \, F_{\mu \nu}F^{\mu \nu}+\frac{k}{4\pi}\,
\epsilon ^{\mu \nu \lambda}A_{\mu}\partial _{\nu}A_{\lambda}, \nonumber \\
F_{\mu \nu} & = & \partial _{\mu}A_{\nu}-\partial _{\nu}A_{\mu}.
\end{eqnarray}

The Hamiltonian and Poisson brackets (PB's) for (2.1) are
\begin{eqnarray}
H & = & \int d^{2}x \: {\cal H}, \nonumber \\
{\cal H} & = & \frac{e^{2}}{2} [(\Pi _{i} +\frac{k}{4\pi}\epsilon
_{ij}A_{j})^{2} + \frac{1}{e^{4}}(\epsilon _{ij}\partial _{i}A_{j})^{2}]
\label{2.2}
\end{eqnarray}
\begin{eqnarray}
\{ A_{i}(x),A_{j}(y)\} & = & \{ \Pi _{i}(x),\Pi _{j}(y)\} =0, \nonumber \\
\{ A_{i}(x),\Pi _{j}(y)\} & = & \delta _{ij}\delta ^{2}(x-y). \label{2.3}
\end{eqnarray}
Here and in what follows, all fields are to be evaluated at some fixed
time while the PB's are at equal times.  Thus $x^{0}=y^{0}$ in (\ref{2.3})\@.
Also $A$ and $\Pi$ give the magnetic field $B$ and the components $E_{i}$ of
the electric field by the formulae
\begin{eqnarray}
B & = & \epsilon _{ij} \partial _{i}A_{j}, \nonumber \\
E_{i} & = & e^{2}(\Pi _{i}+\frac{k}{4\pi}\epsilon _{ij}A_{j}). \label{2.4}
\end{eqnarray}

Dynamics is not entirely given by (2.2,3) as it must be supplemented by
the Gauss law
\begin{equation}
\frac{1}{e^{2}}\partial _{i}E_{i}- \frac{k}{2\pi}\epsilon _{ij}
\partial _{i}A_{j}=\partial _{i}\Pi _{i}-\frac{k}{4\pi}\epsilon _{ij}\partial
_{i}A_{j}\approx 0,         \label{2.5}
\end{equation}
$\approx$ denoting weak equality in the sense of Dirac \cite{11}.
Note that time evolution under Hamiltonian $H$ above generates no new
constraint from (\ref{2.5}).

As discussed elsewhere \cite{5}, in order that the Gauss law generates
canonical
transformations, it needs to be differentiable in the canonical
variables.  This means that the correct expression for Gauss law
is not really (\ref{2.5}), but
\begin{equation}
{\cal G}(\Lambda ^{(0)}):=-\int _{D}d^{2}x\,\partial _{i}\Lambda ^{(0)}[\Pi
_{i}-\frac{k}{4\pi}\epsilon_{ij}A_{j}]\approx 0 \label{2.6}
\end{equation}
where the test functions $\Lambda ^{(0)}$ vanish on $\partial D$:
\begin{equation}
\Lambda ^{(0)} |_{\partial D}=0   \label{2.7}
\end{equation}
{[ Here and elsewhere, the superscript denotes the degree of the form.
Thus $\Lambda ^{(0)}$ is a zero form or a function.]}
Because of (\ref{2.7}), (\ref{2.6}) follows from (\ref{2.5}):
\begin{equation}
0  \approx \int _{D} d^{2}x\,\Lambda ^{(0)} \partial _{i}(\Pi _{i}-
\frac{k}{4\pi}\epsilon_{ij}
A_{j})={\cal G}(\Lambda ^{(0)}) + \int _{\partial D}\Lambda ^{(0)}(\Pi _{i}-
\frac{k}{4\pi}\epsilon_{ij}A_{j})\epsilon_{ik}dx^{k}
={\cal G}(\Lambda ^{(0)} ) .\label{2.8}
\end{equation}
Our task is to quantize (\ref{2.2},3) subject to condition (\ref{2.6}).

In this paper, we will freely use the notation of differential forms and
write $A_{i}, \Pi _{i}, E_{i}$ and $F_{ij}$ in terms of the one and two forms
$A, \Pi , E$ and $F$.  The latter are defined by
\begin{eqnarray}
\alpha =\alpha _{i}dx^{i} & \mbox{for} & \alpha =A, \Pi \ \mbox{or}\
E, \nonumber \\
\vspace{3mm}
F=dA & = & \frac{1}{2}\,F_{ij}\,dx^{i}dx^{j}.      \label{2.9}
\end{eqnarray}
Wedge symbols between differential forms will be entirely omitted.  The
Hodge star operation \cite{10} will be denoted by $*$.  Thus
\begin{eqnarray}
*\alpha =\epsilon _{ij}\alpha ^{j}dx^{i}, &  &  *F=\epsilon _{ij}\partial
^{i}A^{j}=B,\nonumber \\
\
**\alpha =-\alpha , & &  **F=F,\ \ **B=B.     \label{2.10}
\end{eqnarray}

In Hodge theory \cite{10}, one introduces a Hilbert space for forms of degree
$p$
with scalar product
\begin{equation}
(\alpha ^{(p)},\beta ^{(p)})=(-1)^{p}\int \overline{\alpha}^{(p)}*\beta
^{(p)}\label{2.11}
\end{equation}
where the bar denotes complex conjugation.
For $p=1$, therefore,
\begin{eqnarray}
(\alpha ^{(1)},\beta ^{(1)}) & = & \int \overline{\alpha}_{i}^{(1)}\beta
_{i}^{(1)}d^{2}x, \nonumber \\
\alpha ^{(1)}=\alpha _{i}^{(1)}dx^{i}, & &
\beta ^{(1)}=\beta _{i}^{(1)}dx^{i}.   \label{2.12}
\end{eqnarray}
We shall adopt this scalar product hereafter.

In the notation of (\ref{2.12}), $H$ and ${\cal G}(\Lambda ^{(0)})$ can be
written as
\begin{equation}
H=\frac{e^{2}}{2}[(\Pi +\frac{k}{4\pi}*A, \Pi +\frac{k}{4\pi}*A)
+\frac{1}{e^{4}}(A, *d*dA)],      \label{2.13}
\end{equation}
\begin{equation}
{\cal G}(\Lambda ^{(0)})=(d\Lambda ^{(0)},-\Pi +\frac{k}{4\pi}*A)\approx 0.
\label{2.14}
\end{equation}

For quantization, at least in our approach based on mode expansions of one
forms, it is important to write the Gauss law as the scalar product between
one forms as in (\ref{2.14})\@.  We can do so because the surface term in
(\ref{2.8}) vanishes by (\ref{2.7})\@.  We can in fact justify (\ref{2.7}) as a
condition necessary for quantization as the latter requires the equivalence
of (\ref{2.6}) and (\ref{2.5})\@.  For the same reason, it
is equally important for us to write $H$ as well in terms of scalar products
of one forms.  Now (\ref{2.13}) follows from (\ref{2.2}) on partial integration
only if
certain surface terms are discarded, and they are not obviously zero.  For
the boundary conditions (BC's) to be derived below, they vanish only for
$\lambda =0$ or for $\lambda =\infty$, $\lambda$ being the parameter referred
to in the Introduction.  We will argue elsewhere \cite{6}
that (\ref{2.13}) is nevertheless correct if $D$ is regarded as surrounded
for example by a superconductor, the edge being an idealization of the
transition region from the interior of $D$ to its exterior.  In that case,
the exterior as well leads to boundary terms which cancel the ones mentioned
above.  Incidentally, \cite{6} will also establish the BC's by regarding
$D$ as surrounded by a superconductor, $1/\lambda$ being its
penetration
depth.  In any case, we will assume the validity of (\ref{2.13}) hereafter.

Our strategy for quantization is a conventional one.  It consists of expanding
$A$ and $\Pi$ in the complete set of eigenfunctions of the operator $*d*d$.
For $k=0$, it is then readily seen to give an expression for $H$ in
terms of modes which can be arranged to have the commutators of an infinite
number of independent creation-annihilation operators.  As the Gauss law too
assumes a simple form, we can in this way hope to quantize the Lagrangian.  In
case $\lambda$ is zero, the
situation is not much more complicated for $k\neq 0$ if we can find the
eigenstates and eigenvalues of $*d*d$, that being also the more difficult part
of the calculation for $k=0$.  This operator seems to have a central role also
when both $k$ and $\lambda$ differ from zero.  We shall see this in Section 4.

We have therefore to study $*d*d$ and in particular to find a domain [or BC's]
for it so that it is self-adjoint \cite{12}.  We will
study only local BC's which mix eigenfunctions and their
derivatives at a given point.  We will require the BC's to be
compatible with the conservation of angular momentum.  We will also require
that the energy spectrum of
the second quantized Hamiltonian has a lower bound.  With these restrictions,
the BC's depend on a real nonnegative parameter $\lambda$.  The domain defined
by these BC's will be denoted
by ${\cal D}_{\lambda}$.

The property defining a domain $\cal D$ of a self-adjoint operator $S$ is the
following \cite{12}:
\newline The expression \mbox{$(\chi ,S\varphi )-(S\chi ,\varphi )$}
vanishes for all $\varphi \in {\cal D}$ iff $\chi \in {\cal D}.$

In our problem,
\begin{equation}
({\cal B}^{(1)}, *d*d{\cal A}^{(1)})-(*d*d{\cal B}^{(1)}, {\cal A}^{(1)})
=\int _{\partial D}[*d\bar{\cal B}^{(1)}{\cal A}^{(1)} -\bar{\cal B}^{(1)}
*d{\cal A}^{(1)}], \label{2.15}
\end{equation}
${\cal A}^{(1)}$ and ${\cal B}^{(1)}$ being one forms.  Using the preceding
remark, and imposing also locality and angular momentum conservation [but no
condition on energy yet], we find
for the domain,
\begin{eqnarray}
\hat{\cal D}_{\lambda} & = & \{ {\cal A} ^{(1)}|\ *d{\cal A} ^{(1)}=-\lambda
{\cal A} ^{(1)}_{\theta} \mbox{\ for\ }  |\vec{x}|=R\ ,\ \lambda \in
{\bf R}^{1} \},
\nonumber \\
{\cal A} ^{(1)}_{\theta} & := & {\cal A} ^{(1)}_{i} \frac{1}{r}
 \frac{\partial x^{i}}
{\partial \theta}(r,\theta),                  \label{2.16}
\end{eqnarray}
$r, \theta$ being polar coordinates on $D$ with $r=R$ giving its boundary.
{[ $\lambda$ can be a non-trivial function of $\theta$ if angular
momentum conservation is not demanded.\@]}
The members of $\hat{\cal D}_{\lambda}$ are also of course required to be
square integrable.  It is easy to check that $\hat{\cal D}_{\lambda}$ fulfills
the criterion stated above for defining a self adjoint operator.  Note that
$\lambda$ can be negative for
$\hat{\cal D}_{\lambda}$.

As can be seen from (\ref{2.13}), the Hamiltonian will have a lower bound only
if $*d*d$ has no negative eigenvalue.  It is
the requirement that this bound exists which leads to the condition
$\lambda \geq 0$.  We can show this as follows.  Consider \mbox{$
({\cal A}^{(1)},*d*d{\cal A}^{(1)})$} for ${\cal A}^{(1)}\in \hat{\cal D}_
{\lambda}$.  It can be written according to
\begin{eqnarray}
({\cal A}^{(1)},*d*d{\cal A}^{(1)}) & = & \ (d{\cal A}^{(1)},d{\cal A}^{(1)})
-\int _{\partial D}\bar{\cal A}^{(1)}(*d{\cal A}^{(1)})\nonumber \\
& = & (d{\cal A}^{(1)}, d{\cal A}^{(1)}) + \lambda \int _{\partial
D}|{\cal A}^{(1)}_{\theta}|^{2} Rd\theta  \; . \label{2.17}
\end{eqnarray}
This expression is nonnegative for all ${\cal A}^{(1)}$ if and only if
\begin{equation}
\lambda \geq 0                                        \label{2.18}
\end{equation}
thereby establishing the result.  Our domain is thus
\begin{equation}
{\cal D}_{\lambda} = \hat{\cal D}_{\lambda} \; \mbox{\ for } \; \lambda \geq 0
\; .
\end{equation}

The interpretation of $1/\lambda$ as the
penetration depth requires the nonnegativity
of $\lambda$.  It is striking that the physical criterion $H\geq 0$ also
leads to the same condition on $\lambda$.

\sxn{Degenerate Harmonic Modes: Classical Theory}

In Sections 3 and 4, we will assume the zero value for $\lambda$ which
corresponds to infinite penetration depth \cite{6}. We can then regard the
disk and its
exterior as both describing normal (and not superconducting) media.

In this Section, we will classically establish that there are infinitely many
modes with exactly the same frequency $\frac{e^{2}}{2\pi}|k|,\;\;k\neq 0$\@.
They lead to
infinite degeneracy for the energy $E_{H}=\frac{e^{2}|k|}{2\pi}$ in quantum
theory.  Although the
frequencies of these modes approach zero as $k\rightarrow 0$, we will also
point out
that there is no limitation on their energies. We will see in Section 6 that
they behave like momenta in this limit.  The presence of infinitely many of
these modes
also leads to the infinite degeneracy of their energies in the
$k=0$ quantum theory.

The field equations for the Lagrangian $L$ are
\be
\frac{1}{e^{2}} \, \partial_{0}E + \frac{1}{e^{2}}\,*dB -
\frac{k}{2\pi}\,*E = 0 \;, \label{3.1}
\ee
\be
\frac{1}{e^{2}}\,d*E + \frac{k}{2\pi}\,F = 0 \; . \label{3.2}
\ee

The correct interpretation of (\ref{3.1}), compatible with quantization, is
obtained by smearing it with test forms $\Lambda ^{(1)}\in {\cal D} _{0}$
\cite{5} :
\be
\frac{1}{e^{2}}\,(\Lambda ^{(1)},\partial_{0}E) +
\frac{1}{e^{2}}\,(\Lambda ^{(1)},*dB) - \frac{k}{2\pi}\,(\Lambda ^{(1)},*E)=0
\;.\label{3.3}
\ee
As for (\ref{3.2}), it should be read as (\ref{2.14}) with $d\Lambda ^{(0)} \in
{\cal D}_{0}$ and $\Lambda^{(0)}$ fulfilling (\ref{2.7}).

If
\be
z = x_{1} + i \: x_{2} \; ,               \label{3.4}
\ee
the functions given by
\be
z^{n} \;\;, \;\; \bar{z}^{n} \;\;\;\;  n = 1,2,3,\ldots     \label{3.5}
\ee
are harmonic. The one forms $h_{n}^{(1)}$, $\bar{h}_{n}^{(1)}$ defined by
\be
h_{n}^{(1)}(x) = \frac{1}{\sqrt{2\pi n}\, R^{n}}dz^{n} \;\;, \;\;
\bar{h}_{n}^{(1)}(x)
=\frac{1}{\sqrt{2\pi n}\, R^{n}} d\bar{z}^{n}, \;\;\;\; n= 1,2,3, \ldots
\label{3.6}
\ee
belong to ${\cal D}_{0}$ and are null states of $*d*d$ :
\be
 *d*dh_{n}^{(1)} = *d*d\bar{h}_{n}^{(1)} = 0 \; .     \label{3.7}
\ee
They are also orthonormal,
\begin{eqnarray}
(h_{n}^{(1)},h_{m}^{(1)}) & =\delta_{nm}= & (\bar{h}_{n}^{(1)},
\bar{h}_{m}^{(1)}), \nonumber\\
(h_{n}^{(1)},\bar{h}_{m}^{(1)}) & = & 0      ,  \label{3.8}
\end{eqnarray}
and are eigenstates of $*$:
\be
 *h_{n}^{(1)}=ih_{n}^{(1)}\;\; , \;\; *\bar{h}_{n}^{(1)}=-i\bar{h}_{n}^{(1)}.
\label{3.85}
\ee

The scalar products of $*dB$ with $h_{n}^{(1)}$ and $\bar{h}_{n}^{(1)}$ are
zero
by the BC's.  For example,
\be
(h_{n}^{(1)},*dB) = \int_{D}\frac{1}{\sqrt{2\pi n}\, R^{n}} d\bar{z}^{n}\:
dB = - \int_{\partial D}\frac{1}{\sqrt{2\pi n}\, R^{n}} d\bar{z}^{n}\:B = 0
\label{3.9}
\ee
since \[ B|_{\partial D} = *dA|_{\partial D}.\]

The ``harmonic'' modes of $E$ are
\be
(h_{n}^{(1)},E)\;\;\;\;  \mbox{and} \;\;\;\; (\bar{h}_{n}^{(1)},E).
\label{3.95}
\ee
According to (\ref{3.1}),
\begin{eqnarray}
\partial_{0} (h_{n}^{(1)},E) & = & i\frac{e^{2}k}{2\pi}\,(h_{n}^{(1)},E) \; ,
\nonumber \\
\partial_{0} (\bar{h}_{n}^{(1)},E) & = & -i\frac{e^{2}k}{2\pi}\,
(\bar{h}_{n}^{(1)},E) \; .    \label{3.10}
\end{eqnarray}
Also ${\cal G}(z^{n})$ and ${\cal G}(\bar{z}^{n})$ need not vanish,
and Gauss law
places no condition on these modes, because $z^{n}$ and $\bar{z}^{n}$ do not
vanish on $\partial D$.

Equation (\ref{3.10}) shows the presence of infinitely many modes, all with the
same frequency $e^{2}|k|/2\pi$ as claimed.

All these modes become constants of motion in the limit $k \rightarrow 0$.
Although their frequencies are zero in
this limit, that is not the case with their energies. One can see this from the
fact that the Hamiltonian contains a term proportional to the integral of
$E^{2}$ and hence receives contributions from $(h_{n}^{(1)},E)$ and
$(\bar{h}_{n}^{(1)},E)$. We will further discuss these modes and their effect
on energy for $k=0$ in Section 6.

\sxn{Quantization for Nonzero $k$ and Zero $\lambda$}

{\bf 4.1. The Spectrum and Eigenfunctions of $*d*d$\@.}\\
\par
The first task in quantization is to investigate the eigenvalue problem for
$*d*d$ which we now carry out.

The eigenvalue equation for $*d*d$ is
\be
*d*d \Lambda _{n}^{(1)} = \omega_{n}^{2} \, \Lambda _{n}^{(1)}\; , \; \Lambda
_{n}^{(1)} \in {\cal D}_{0}\;\;\; \mbox{or} \;\;\; *d\Lambda _{n}^{(1)}|
_{\partial D}=0 \label{4.1}
\ee
where, as shown previously, $\omega_{n}^{2} \geq 0$ .[In our conventions, zero
is also regarded as a permissible eigenvalue.]

Consider first the null modes $\xi _{n}^{(1)}$ with $\omega_{n}^{2} = 0\,$:
\begin{equation}
*d*d\xi _{n}^{(1)} = 0 \; . \label{4.1a}
\end{equation}
Then $d*d\xi _{n}^{(1)}$ is zero and $*d\xi_{n}^{(1)}$ is a constant.
This constant must be zero by the BC's so that
$d\xi _{n}^{(1)} = 0$ or $\xi _{n}^{(1)}$ is closed.  For $D$, this implies
that $\xi _{n}^{(1)}$ is exact:
\be
 \xi _{n}^{(1)} = d\xi _{n}^{(0)} .      \label{4.2}
\ee

There are two cases to be considered:

a) $\xi _{n}^{(0)}|_{\partial D} \neq 0 \,$, and

b) $\xi _{n}^{(0)}|_{\partial D} = 0 $.

In the former case, we can show that $\xi _{n}^{(0)}$ are all given by the
harmonic functions $z^{n}$ and $\bar{z}^{n}$ \cite{10}. For suppose that
$d\eta_{m}^{(0)}$ is an eigenfunction [or an ``eigenform", although we will not
use that phrase hereafter] orthogonal to all $h_{n}^{(1)}$ and
$\bar{h}_{n}^{(1)}$:
\be
(d \eta_{m}^{(0)},h_{n}^{(1)}) = (d \eta_{m}^{(0)},\bar{h}_{n}^{(1)}) = 0.
\label{4.3}
\ee
Using (\ref{3.85}) and (\ref{3.6}), we can write (\ref{4.3}) as
\begin{eqnarray}
(d \eta_{m}^{(0)},h_{n}^{(1)}) & = & -i\int_{D} d \bar{\eta}_{m}^{(0)}\,
h_{n}^{(1)}=  -i\int_{\partial D} \bar{\eta}_{m}^{(0)}\,
\frac{1}{\sqrt{2\pi n}\,R^{n}}dz^{n}=0 \; , \nonumber \\
(d\eta_{m}^{(0)},\bar{h}_{n}^{(1)}) & = & i\int_{D} d\bar{\eta}_{m}^{(0)}\,
\bar{h}_{n}^{(1)}=  i\int_{\partial D} \bar{\eta}_{m}^{(0)}\,
\frac{1}{\sqrt{2\pi n}\, R^{n}}d\bar{z}^{n}  =  0 \; .  \label{4.4}
\end{eqnarray}
Hence
\be
\int_{\partial D} \bar{\eta}_{m}^{(0)} \, e^{in\theta} \, d\theta = 0 \;
\mbox{for}\;\; n = \pm 1, \pm 2, \ldots       \label{4.5}
\ee
or $\eta_{m}^{(0)}$ is a constant $c_{m}$ on $\partial D$\@. As the
eigenfunction is the one form $d\eta_{m}^{(0)} = d(\eta_{m}^{(0)} - c_{m})\,$,
we can replace $\eta_{m}^{(0)}$ by $\tilde{\eta}_{m}^{(0)} = \eta_{m}^{(0)} -
c_{m}$ which fulfills $\tilde{\eta}_{m}^{(0)}|_{\partial D} = 0\,$.
Calling $\tilde{\eta}_{m}^{(0)}$ again as $\eta_{m}^{(0)}$, we thus have the
condition
\be
\eta_{m}^{(0)}|_{\partial D} = 0\,.  \label{4.6}
\ee
In other words, for case $a$, all $\xi _{n}^{0}$ are given by harmonic modes.
[We use the phrase ``harmonic modes" interchangeably to denote
$h_{n}^{(1)},\bar{h}_{n}^{(1)}$ and the scalar product of one forms with
$h_{n}^{(1)},\bar{h}_{n}^{(1)}$. In the same way, we will use the phrase
``mode" to denote any eigenfunction and also its scalar product with a one
form.]

Now $\eta_{m}^{(0)}$ can not be harmonic unless it is identically zero. For if
$\eta_{m}^{(0)}$ is harmonic,
\be
0 = (\eta_{m}^{(0)},*d*d\eta_{m}^{(0)}) = \int_{D} |\partial_{i}
\eta_{m}^{(0)}|^{2} \: d^{2}x        \label{4.7}
\ee
or $\eta_{m}^{(0)}$ is a constant which is zero by our boundary condition.

The null eigenvectors are thus given by the following:
\begin{eqnarray}
{}~ & a) & \;\;\;\;\;h_{n}^{(1)}(x)= \frac{1}{\sqrt{2\pi n}R^{n}}dz^{n}\;\;
,\;\;
\bar{h}_{n}^{(1)}(x)=\frac{1}{\sqrt{2\pi n}R^{n}}d\bar{z}^{n},\;\; n=1, 2,
\ldots \hspace{0.5cm} , \label{4.8} \\
{}~ & b) & \;\;\;\;\;d\eta_{n}^{(0)},\;\; \eta_{n}^{(0)}|_{\partial D}=0 \;\;
,\;\; \eta_{n}^{(0)} \mbox{\ is not harmonic}. \label{4.9}
\end{eqnarray}
We recall also that the modes $h_{n}^{(1)},\; \bar{h}_{n}^{(1)}$ are
eigenstates
of $*$ :
\begin{flushright}
$*h_{n}^{(1)}=ih_{n}^{(1)}\;\; , \;\;
*\bar{h}_{n}^{(1)}=-i\bar{h}_{n}^{(1)}$ . \hspace{5cm} (3.9)
\end{flushright}

Let $\hat{\Psi} _{n}^{(1)}$ be an eigenfunction for $\omega _{n}^{2}\neq 0$:
\be
 *d*d\hat{\Psi} _{n}^{(1)} = \omega _{n}^{2}\hat{\Psi} _{n}^{(1)}, \;\;\;
\omega
_{n}^{2}\neq 0. \label{4.10}
\ee
We will next show that 1) $ *\hat{\Psi} _{n}^{(1)}$ is a type $b$
eigenfunction, and
2) every type $b$ eigenfunction can be obtained as a superposition of
$*\hat{\Psi} _{n}^{(1)}$ using solutions of (\ref{4.10}). We can
thus construct all type $b$ eigenfunctions from solutions of (\ref{4.10}).

By (\ref{4.10}),
\be
 *d*d (*\hat{\Psi} _{n}^{(1)}) = -\frac{1}{\omega _{n}^{2}}(*d*d)(d*d\hat{\Psi}
_{n}^{(1)})
= 0. \label{4.11}
\ee
So $*\hat{\Psi} _{n}^{(1)}$ is an eigenfunction for zero eigenvalue.

Now $\hat{\Psi} _{n}^{(1)}$, which has $\omega _{n}^{2}\neq 0$ is orthogonal to
all
$h_{m}^{(1)}, \bar{h}_{m}^{(1)}$ which correspond to zero eigenvalues.  Hence,
in view of (\ref{3.85}),
\begin{eqnarray}
(\hat{\Psi} _{n}^{(1)}, h_{m}^{(1)}) & =0= & (*\hat{\Psi} _{n}^{(1)},
h_{m}^{1}), \nonumber \\
(\hat{\Psi} _{n}^{(1)}, \bar{h}_{m}^{(1)}) & =0= & (*\hat{\Psi} _{n}^{(1)},
\bar{h}_{m}
^{(1)}). \label{4.12}
\end{eqnarray}
It follows that $*\hat{\Psi} _{n}^{(1)}$ is of type $b$.

In order to show item 2 above, let
\be
 *d\hat{\Psi} _{n}^{(1)} = F_{n}^{(0)}. \label{4.13}
\ee
Then by (\ref{4.10}),
\begin{eqnarray}
\hat{\Psi} _{n}^{(1)} & = & \frac{1}{\omega _{n}^{2}} *dF_{n}^{(0)},
\label{4.14}\\
{*}\hat{\Psi} _{n}^{(1)} & = & -\frac{1}{\omega _{n}^{2}}
dF_{n}^{(0)},\label{4.15}
\end{eqnarray}
so that $\hat{\Psi} _{n}^{(1)}$ and its $*$ are given by $F_{n}^{(0)}$\@.
$F_{n}^{(0)}$ fulfills the differential equation and BC
\be
-\Delta F_{n}^{(0)}= *d*dF_{n}^{(0)} = \omega _{n}^{2} F_{n}^{(0)},
\label{4.16}
\ee
\be
 F_{n}^{(0)}|_{\partial D} = 0 \label{4.17}
\ee
as follows from (\ref{4.14}), (\ref{4.13}).  $\Delta$ here is the Laplacian.

Now if a type $b$ form $d\eta ^{(0)}$ is orthogonal to all $*\hat{\Psi}
_{n}^{(1)}$, then on partial integration, we get
\be
\int \bar{F}_{n}^{(0)}d*d\eta ^{(0)} =0,\;\;\;\forall n. \label{4.18}
\ee
But (\ref{4.16}) is well known to have a complete set of eigenfunctions
for the BC (\ref{4.17}) in the Hilbert space with scalar product (\ref{2.11}).
Hence $\eta ^{(0)}$ is harmonic and by (\ref{4.9}), zero.  This shows item 2.

It remains to solve for $F_{n}^{(0)}$.  We set
\begin{eqnarray}
n & = & NM , \nonumber \\
F_{n}^{(0)}(x) \equiv F_{NM}^{(0)}(x) & = & e^{iN\theta}G_{NM}^{(0)}(r),
\;\;\; N\in {\bf Z} \equiv \{0,\pm 1,\pm 2,\ldots\} \label{4.19}
\end{eqnarray}
and find that (\ref{4.16}) becomes Bessel's equation for $G_{NM}^{(0)}$\@:
\be
\{ \frac{d^{2}}{dr^{2}} + \frac{1}{r}\frac{d}{dr} + (\omega _{NM}^{2} -
\frac{N^{2}}{r^{2}})\} G_{NM}^{(0)}(r) = 0. \label{4.20}
\ee
We have written $\omega _{n}^{2}$ as $\omega _{NM}^{2}$ in (\ref{4.20}).
The solution of (\ref{4.20}) regular at $r=0$ is $J_{N}(\omega _{NM}r)$.
Using also (\ref{4.17}), we thus have
\begin{eqnarray}
F_{NM}^{(0)}(x) & = & e^{iN\theta}G_{NM}^{(0)}(r),\nonumber \\
G_{NM}^{(0)}(r) & = & J_{N}(\omega _{NM}r), \label{4.21}
\end{eqnarray}
\be
J_{N}(\omega _{NM}R) \;\; = \;\; 0. \label{4.22}
\ee
Note that $G_{NM}^{(0)}$ vanishes identically if $\omega _{NM}=0$ and $N\neq 0$
while $\omega_{0M}=0$ is not a solution of (\ref{4.22}) for $N=0$\@.  Therefore
$\omega _{NM}^{2}\neq 0$ in (\ref{4.21}).  Also solutions for $\omega _{NM}<0$
are linearly dependent on those for $\omega _{NM}>0$ since $J_{N}(-x)=(-1)^{N}
J_{N}(x)$.  We hence assume that
\be
\omega _{NM} > 0. \label{4.23}
\ee

Now the positive roots of $J_{N}(x)$ form an infinite countable set \cite{13}.
Thus the set $\{1,2,3, \ldots \}$ for $M$ is adequate to label the
eigenfunctions for a given $N$\@.  We also choose $M$ so that
\be
0\;<\;\omega _{N1}\;<\;\omega _{N2}\;<\; \ldots \hspace{0.5cm} . \label{4.24}
\ee
Note that since $\omega _{NM}^{2}\neq \omega _{NM'}^{2}$ if $M\neq M'$, we
have the orthogonality relation
\be
\int _{0}^{R}dr\,r\,\bar{G}_{NM}^{(0)}(r)\,G_{NM'}^{(0)}(r)\;=0 \;\;
\mbox{if} \;\; M \neq M'. \label{4.25}
\ee

Although $N$ can be any integer in (\ref{4.20}), the differential
operator there with $\omega_{NM}^{2}$ as eigenvalue depends only on $N^{2}$.
It is therefore the case that
\be
\omega _{-NM} =\omega _{NM}. \label{4.26}
\ee
So we set $G_{-NM}^{(0)}=G_{NM}^{(0)}$\@.  Also $G_{NM}^{(0)}$ will be chosen
to be real, (\ref{4.20}) allowing this choice.  We get in this way the
equations
\be
G_{-NM}^{(0)}=\bar{G}_{NM}^{(0)}=G_{NM}^{(0)} \; , \label{4.27}
\ee
\be
F_{-NM}^{(0)}=\bar{F}_{NM}^{(0)} \; . \label{4.28}
\ee
Replacing $NM$ by $nm$ and setting
\be
 *\Psi _{nm}^{(1)}=N_{nm}\,dF_{nm}^{(0)},\;\; n\in {\bf Z} ,\;\; m\in
\{ 1,\;2,\;3,\ldots\} , \label{4.29}
\ee
where the constant $N_{nm}$ is fixed by
\be
(*\Psi _{nm}^{(1)},\; *\Psi _{nm}^{(1)}) = 1,\;\; N_{nm}\; > \; 0, \label{4.30}
\ee
it follows that
\be
\overline{\Psi _{nm}^{(1)}} =\Psi _{-nm}^{(1)},\;\;\overline{*\Psi _{nm}^{(1)}}
 = *\Psi _{-nm}^{(1)} \label{4.31}
\ee
and that
\be
\Psi _{nm}^{(1)},\;\; *\Psi _{nm}^{(1)},\;\; h_{n}^{(1)},\;\;
\bar{h}_{n}^{(1)} \label{4.32}
\ee
form an orthonormal and complete set of eigenfunctions, the first for $\omega
_{nm}^{2}\neq 0$, and the rest for zero eigenvalue, (\ref{4.22}) giving the
nonzero eigenvalues.

A mathematical point may be noted here: it is not true that $ *d\eta ^{(0)}\in
{\cal D}_{0}$ for all $\eta ^{(0)}$ vanishing in $\partial D$\@.
For a generic $\eta ^{(0)}$ with
$\eta ^{(0)}|_{\partial D}=0,\;\; *d\eta ^{(0)}$ will not fulfill the
boundary condition appropriate for elements of ${\cal D}_{0}$\@.  Nevertheless
there exists a complete set $ dF_{nm}^{(0)}$ for type $b$ eigenfunctions
with the properties $F_{nm}^{(0)}|_{\partial D}=0$ and $*dF_{nm}^{(0)}
\in {\cal D}_{0}$.\\

{\bf 4.2 The Hamiltonian and its Fock Space}\\
\par
The first calculations we do for the purpose of constructing the Hamiltonian
and its Fock space concern the mode expansions of fields and the derivations of
the Gauss law in terms of these modes.  Later, we will find a suitable Fock
space for the Hamiltonian, and also
comment on its spectrum and in particular on its remarkable degeneracies.

The mode expansions of $A$ and $\Pi$ are the following if the convention of
summing over repeated indices is adopted:
\begin{eqnarray}
A & = & a_{nm}\Psi _{nm}^{(1)} + a_{nm}^{(*)}*\Psi _{nm}^{(1)} + \alpha _{n}
h_{n}^{(1)} +\alpha _{n}^{\dagger}\,\bar{h}_{n}^{(1)},\nonumber \\
\Pi & = & \pi _{nm}\Psi _{nm}^{(1)} + \pi _{nm}^{(*)}*\Psi _{nm}^{(1)} +
p_{n}h_{n}^{(1)} + p_{n}^{\dagger}\,\bar{h}_{n}^{(1)}. \label{4.33}
\end{eqnarray}
{[$n$ is summed over all integers in the first two terms of $A$ and $\Pi$, and
over positive integers in the last two terms.  $m$ is summed over positive
integers only.]}
The reality of $A$ and $\Pi$ has been partially used in (\ref{4.33}).  Also,
in view of
this reality and (\ref{4.31}), there are the further relations
\be
\chi _{nm}^{\dagger} =\chi _{-nm} \; \; \mbox{for} \; \; \chi = a,\; a^{(*)},\;
\pi ,\; \pi ^{(*)} \; . \label{4.34}
\ee
It is also sometimes convenient to define $\alpha _{-n}, p_{-n}$ and their
adjoints for $n\geq 0$ by
\be
\gamma _{n}^{\dagger} =\gamma _{-n} \mbox{ for } \gamma =\alpha
,p.\label{4.345}
\ee

The PB's (\ref{2.3}) give the commutation relations (CR's) for the modes.
All the nonzero commutators are given by
\begin{eqnarray}
{[a_{nm}\; ,\; \pi _{n'm'}]} & =i \delta _{n+n',0} \, \delta _{mm'}= &
{[a_{nm}^{(*)}\; ,\; \pi _{n'm'}^{(*)}]} \; ,\nonumber\\
{[a_{nm}\; ,\; \pi _{n'm'}^{\dagger}]} & =i \delta _{nn'} \delta _{mm'}= &
{[a_{nm}^{(*)}\; ,\; \pi _{n'm'}^{(*)\dagger}]} \; ,\nonumber\\
{[\alpha _{n}\; ,\; p_{m}^{\dagger}]} & =  i \delta _{nm} = &
{[\alpha _{n}^{\dagger} \; ,\; p_{m}]} \; ,\nonumber\\
{[\alpha _{n}\; ,\; p_{m}]} & = i\delta _{n+m,0} = & {[\alpha _{n}^{\dagger}\;
,\; p_{m}^{\dagger}]}\; .  \label{4.35}
\end{eqnarray}

The expression for the Gauss law in terms of the modes above follows from
(\ref{2.14}) by substituting $N_{nm}\; dF_{nm}$ for $d\Lambda ^{(0)}$.  If
$|\cdot \rangle$ is a physical state, the Gauss law in quantum theory
reads
\be
{\cal G}_{nm}|\cdot \rangle \equiv {\cal G}(N_{nm}\;
dF_{nm}^{(0)})|\cdot \rangle = (*\Psi_{nm}^{(1)}\; ,-\Pi +
\frac{k}{4\pi}*A )|\cdot \rangle = (-\pi _{nm}^{(*)} +
\frac{k}{4\pi}a_{nm})|\cdot \rangle =  0 \; . \label{4.36} \ee
Note that $a_{nm}^{(*)}$ do not commute with ${\cal G}_{-nm}$ and hence
are not observable.

It is very convenient to introduce the mode expansion of the electric field for
studying the Hamiltonian.  Let us adopt the summation conventions involved in
(\ref{4.33}) hereafter. We can then write
\begin{eqnarray}
{}~ & ~ & E = e_{nm}\Psi _{nm}^{(1)} + e_{nm}^{(*)}*\Psi _{nm}^{(1)} + c _{n}
h_{n}^{(1)} + c_{n}^{\dagger} \, \bar{h}_{n}^{(1)}, \nonumber \\
{}~ & ~ & e_{nm}^{\dag} =  e_{-nm},\;\;\; e_{nm}^{(*)\dagger}\;\; = \;\;
e_{-nm}^{(*)} , \label{4.37}
\end{eqnarray}
where in view of (\ref{2.4}),
\begin{eqnarray}
e_{nm} & = & e^{2}(\pi_{nm} -\frac{k}{4\pi} a_{nm}^{(*)}) \nonumber \\
e_{nm}^{(*)} & = & e^{2}(\pi_{nm}^{(*)}+\frac{k}{4\pi} a_{nm}) \nonumber \\
c_{n} & = & e^{2}( p_{n} +\frac{ik}{4\pi} \alpha_{n}) \nonumber \\
c_{n}^{\dagger} & = & e^{2}(p_{n}^{\dagger} -\frac{ik}{4\pi}
\alpha_{n}^{\dagger}) \label{4.43a}
\end{eqnarray}
The nonzero commutators among these are given by
\begin{eqnarray}
[e_{nm},e_{n'm'}^{(*)}] & = & -i\frac{e^{4}k}{2\pi}\, \delta _{n+n',0} \,
\delta _{mm'}\; ,
\nonumber \\
{[c_{n},c_{m}^{\dagger}]} & = & -\frac{e^{4}k}{2\pi} \, \delta _{nm}\; ,
\label{4.43b}
\end{eqnarray}
as follows from (\ref{4.35}).

The Gauss law (\ref{4.36}) in terms of these modes is
\be
(e_{nm}^{(*)} -\frac{e^{2}k}{2\pi}a_{nm})|\cdot \rangle = 0.
\label{4.385} \ee

The Hamiltonian acting on physical states, after using (\ref{4.385}), is seen
to be
\begin{equation}
H = \frac{1}{2e^{2}}(e_{nm}^{\dagger} \, e_{nm} +[\omega _{nm}^{2} +
(\frac{e^{2}k}{2\pi})^{2}] a_{nm}^{\dagger} \, a_{nm} +
c_{n}^{\dagger} c_{n} + c_{n} c_{n}^{\dagger}) \; .\label{4.44}
\end{equation}

Hereafter for $\lambda=0$, we will work exclusively with $e_{nm}$, $a_{nm}$,
the
harmonic modes $c_{n},p_{n},\alpha_{n}$ and their adjoints. As all of them
commute
with any ${\cal G}_{\rho \sigma}$, and $H$ can be expressed using them alone,
({\ref{4.385}) can and will be ignored hereafter.

The first two sums together in (\ref{4.44}) describe modes which become those
of
the massive photon when $R \rightarrow \infty$. They can be diagonalized
using the operators
\begin{eqnarray}
{\cal A}_{nm} & = &
\frac{1}{\sqrt{2}e}[\frac{e_{nm}}{\sqrt{\Omega_{nm}}}-i\sqrt{\Omega_{nm}}a_{nm}]
\nonumber \\
{\cal A}_{nm}^{\dagger} & = & \frac{1}{\sqrt{2}e}[\frac{e_{nm}^{\dagger}}
{\sqrt{\Omega_{nm}}}+i\sqrt{\Omega_{nm}}a_{nm}^{\dagger}] \nonumber \\
{\cal B}_{nm} & = &
\frac{1}{\sqrt{2}e}[\frac{e_{nm}^{\dagger}}{\sqrt{\Omega_{nm}}}-
i\sqrt{\Omega_{nm}}a_{nm}^{\dagger}] \nonumber \\
{\cal B}_{nm}^{\dagger} & = &
\frac{1}{\sqrt{2}e}[\frac{e_{nm}}{\sqrt{\Omega_{nm}}}+
i\sqrt{\Omega_{nm}}a_{nm}]
\label{4.45} \end{eqnarray}
where
\begin{displaymath}
\Omega _{nm}=[\omega _{nm}^{2}+(\frac{e^{2}k}{2\pi})^{2}]^{1/2}\; \mbox{and} \;
n\geq 0 \; .
\end{displaymath}

These are creation-annihilation operators with commutators
\begin{displaymath}
[\chi _{nm},\chi _{n'm'}] = [\chi _{nm}^{\dagger},\chi _{n'm'}^{\dagger}]
= 0 \; ,
\end{displaymath}
\begin{displaymath}
[\chi _{nm},\chi _{n'm'}^{\dagger}] = \delta_{nn'} \, \delta_{mm'} \; ,
\end{displaymath}
\begin{equation}
\chi ={\cal A} \; \mbox{or} \; \cal B\; . \label{4.46}
\end{equation}

Let us define
\begin{eqnarray}
d_{n} = \frac{1}{e^{2}} \sqrt{\frac{2\pi}{|k|}} c_{n} & , &
d_{n}^{\dagger} = \frac{1}{e^{2}} \sqrt{\frac{2\pi}{|k|}} c_{n}^{\dagger}
\hspace{0.5cm} \mbox{for} \;  k > 0 \; , \nonumber \\
d_{n}^{\dagger} = \frac{1}{e^{2}} \sqrt{\frac{2\pi}{|k|}} c_{n} & , &
d_{n} = \frac{1}{e^{2}} \sqrt{\frac{2\pi}{|k|}} c_{n}^{\dagger}
\hspace{0.5cm} \mbox{for} \;  k < 0 \; , \label{4.46a}
\end{eqnarray}
and introduce the vacuum $|0\rangle$ by
\begin{equation}
\chi _{nm} |0\rangle = d_{n} |0\rangle = 0 \;  ,
\hspace{0.5in} n \geq 0 \: , m \in \{1,2,3,\ldots\} \; .   \label{4.47}
\end{equation}
We also normal order $H$ relative to the vacuum and normalize the energy of
the vacuum to be zero. It then becomes
\begin{equation}
H = \Omega _{nm}({\cal A}_{nm}^{\dagger} \,
{\cal A}_{nm} + {\cal B}_{nm}^{\dagger} \, {\cal B}_{nm}) +
\frac{e^{2}|k|}{2\pi} d_{n}^{\dagger} \, d_{n} \; . \label{4.48}
\end{equation}

The specification of the vacuum using (\ref{4.47}) is incomplete. This is
because
there are observable modes which do not appear in $H$. They are therefore
constants of motion and generate a symmetry group which is known to be the
affine Lie group $\tilde{L}U(1)$. They are the edge observables we have
alluded to before. They are similar to ${\cal G}(\Lambda^{(0)})$ but for the
BC's on $\Lambda^{(0)}$ and are
\begin{displaymath}
q_{n} = (h_{n}^{(1)},-\Pi +\frac{k}{4\pi}*A)=-p_{n}+\frac{ik}{4\pi}
\alpha_{n}
\end{displaymath}
\begin{equation}
q_{n}^{\dagger} = (\bar{h}_{n}^{(1)},-\Pi +\frac{k}{4\pi}*A)=
-p_{n}^{\dagger}-\frac{ik}{4\pi}\alpha_{n}^{\dagger} \; .      \label{4.49}
\end{equation}
Their non-zero commutators are given by
\begin{equation}
[q_{n},q_{m}^{\dagger}] = \frac{k}{2\pi}\delta_{nm} \; .   \label{4.50}
\end{equation}
The specification of $|0\rangle$ thus requires the additional condition
\begin{equation}
Q_{n} \, |0\rangle = 0\; \mbox{ for } n\geq 0 ,        \label{4.50a}
\end{equation}
\begin{eqnarray}
Q_{n} & = & q_{n} \; , \; Q_{n}^{\dagger} = q_{n}^{\dagger} \hspace{0.5cm}
\mbox{for} \; n > 0 \mbox{ and } k > 0 \; , \nonumber \\
Q_{n} & = & q_{n}^{\dagger} \; , \; Q_{n}^{\dagger} = q_{n} \hspace{0.5cm}
\mbox{for} \; n > 0 \mbox{ and } k < 0 \; . \label{4.50b}
\end{eqnarray}

Elsewhere \cite{5}, we have explained why $q_{n}$ and $q_{n}^{\dagger}$ are
called
``edge" observables. Briefly, the reason is as follows. By adding arbitrary
linear
combinations of the Gauss law constraints (\ref{2.6}) we can see that $q_{n}$
for
example is weakly equivalent to any $(d \Lambda^{(0)},-\Pi +\frac{k}{4\pi}*A)$
provided only that $d \Lambda^{(0)}|_{\partial D} = \frac{1}{\sqrt{2\pi
n}R^{n}}dz^{n}|_{\partial D}$. In other words, up to weak equivalence,
$q_{n}$ is
entirely determined by the restriction of $\Lambda^{(0)}$ to the edge $\partial
D$ and hence can be regarded as an edge observable. Similar remarks can be
made about $q_{n}^{\dagger}$ as well.
A better demonstration that these are edge observables will consist in showing
that they commute with all observables localized in the interior of $D$.
We can establish this result by noting that their action on any observable
localized in an open set $U$ in the interior of $D$ is a gauge
transformation for some gauge function $\Lambda$ as seen from their similarity
to ${\cal G}(\Lambda^{(0)})$. Clearly we can deform $\Lambda$ outside $U$ so
that it vanishes on $\partial D$ without affecting this action. So the
action of $q_{n}$, $q_{n}^{\dagger}$ is that of Gauss law operators for
observables localized within $D$. As observables commute with Gauss law
operators, it follows that observables localized in the interior of $D$
commute with $q_{n}$, $q_{n}^{\dagger}$. This is
the result we were after.

Note that the Fock states obtained by
applying polynomials in $Q_{n}^{\dagger}$ ($n>0$) to any state can be
regarded as its
excitations localized at the edge. These excitations cost no energy as $q_{n}$,
$q_{n}^{\dagger}$ do not appear in $H$.

An interesting observation to be noted here is that the angular momentum $L$,
defined as
\[ L = \int d^{2}x\, A_{i}{\cal L}_{v}\Pi _{i}, \mbox{ where } v_{i} =
-\epsilon _{ij}x_{j} \]
and can be expanded using the modes for $\lambda = 0$.  The contribution from
the harmonic modes in this expansion contain factors of the form
\[ l(n)\;(d^{\dagger}_{n}d_{n} -\frac{2\pi}{k}q^{\dagger}_{n}q_{n}) \]
where $l(n)$ is proportional to $n$ and $k$ is the coefficient which appears in
(2.1).

Thus the infinite degeneracy caused because of the absence of $q^{\dagger}_{n}$
and $q_{n}$ in the unperturbed Hamiltonian is removed if an interaction which
couples to the angular momentum is added to the system.\\

\par

{\bf 4.3 The Spectrum, Degeneracies and Symmetries}\\

\par

We will call the state obtained by applying a creation operator once to the
vacuum as a  ``single particle" state, and the corresponding energy as a
``single particle" energy. Energies and degeneracies for the associated
``multiparticle" states, obtained by multiple applications of creation
operators
to the vacuum, follow from those of the single particle states. We will also
freely use phrases like single and multiparticles suggested by these
conventions.

The energies of the single particle states created by ${\cal A}_{nm}^{\dagger}$
and ${\cal B}_{nm}^{\dagger}$ are $\Omega_{nm}$.
These states become the massive photon states as $R \rightarrow \infty$.

The single particles created by $d_{n}^{\dagger}$ have energy
$\frac{e^{2}|k|}{2\pi}$. They are associated
with the modes of Section 3 and are exactly degenerate in energy. Note that
this
energy goes to zero with $k$ and that, in this limit, $c_{n}$ and
$c_{n}^{\dagger}$
become $-e^{2}q_{n}$ and $-e^{2}q_{n}^{\dagger}$.

Following Section 3, the modes of $E$ associated with $h_{n}^{(1)}$,
$\bar{h}_{n}^{(1)}$, such as $c_{n}$ and $c_{n}^{\dagger}$, will be referred to
as
harmonic modes, and the corresponding single particles as harmonic particles.

The Hamiltonian (\ref{4.48}) is invariant under all unitary transformations of
the form
\begin{equation}
d_{n} \rightarrow U_{nn'} \, d_{n'} \; , \hspace{0.25in} d_{n}^{\dagger}
\rightarrow
d_{n'}^{\dagger} \, U_{n'n}^{\dagger} \; , \hspace{0.25in} U^{\dagger} U = 1
\label{4.51}
\end{equation}
and has the infinite constants of motion
\begin{equation}
T_{nm} = d_{n}^{\dagger} d_{m} +d_{m}^{\dagger}d_{n} \; , \hspace{0.5in}
n,m \in \{1,2,3,\ldots\} \label{4.52}
\end{equation}
which can be used to generate these symmetries. No clear interpretation of this
$U(\infty)$ symmetry is known to us.

The single particle states created by $Q_{n}^{\dagger}$ all have zero energy.
The
corresponding multiparticle states too have zero energy. The ground state
energy
of $H$ is thus extremely degenerate.

The symmetry underlying this ground state degeneracy is reasonably well
understood.
The operators $q_{n}$ and $q_{n}^{\dagger}$ commute with $H$ and can be used to
generate this symmetry. In view of (\ref{4.50}), the symmetry group can be
regarded as
the centrally extended loop group $\tilde{L}U(1)$ of $U(1)$. Previous
work shows this symmetry to be a consequence of gauge invariance, a result that
can also be inferred by noting that $q_{n}$, $q_{n}^{\dagger}$ can be obtained
from the Gauss law generator by replacing $d\Lambda^{(0)}$ by  $h_{n}^{(1)},
\bar{h}_{n}^{(1)}$ as remarked previously.

Summarizing, the MCS dynamics on a disk has an enormous
symmetry group and spectral degeneracy when $\lambda =0$. As stated in the
Introduction, the physical meaning of $\lambda$ will be explained elsewhere.

The large symmetry group of MCS dynamics does not survive when the limit $R
\rightarrow \infty$ is taken as can be seen in the following way. This group
depends for its existence on the modes $c_{n}$, $q_{n}$ and their adjoints, and
the definition of the latter involve $h_{n}^{(1)}$ and $\bar{h}_{n}^{(1)}$. But
the latter forms become zero for each fixed $n$ as $R \rightarrow \infty$,
indicating the absence of this group for an infinitely large disk.

\sxn{Quantization for Nonzero $k$ and Positive $\lambda$}

{\bf 5.1. The Spectrum and Eigenfunctions of $*d*d$}\\
\par

The eigenvalue problem for $*d*d$ can be analyzed by straightforward methods.

Among the eigenfunctions in (\ref{4.32}), $*\Psi_{nm}^{(1)}$ continue to be
eigenfunctions with zero eigenvalue for $\lambda>0$ as well. This is because
(a) they are exact and (b) $F_{nm}^{(0)}|_{\partial D}=0$, (a) implying that
$*d*d(*\Psi_{nm}^{(1)})=0$ and (b) implying the satisfaction of the boundary
condition in (\ref{2.16}).

We can find no more one forms in the kernel of $*d*d$. In particular the
harmonic
modes $h_{n}$ and $\bar{h}_{n}$ are not eigenvectors of this operator since
they
do not satisfy the boundary condition in (\ref{2.16}) for $\lambda \neq 0$.

Let
\begin{equation}
*d\hat{\Psi}_{nm}^{(1)} = \hat{F}_{nm}^{(0)} \label{5.1}
\end{equation}
as in (\ref{4.13}), where $\hat{\Psi}_{nm}^{(1)}$ is an eigenfunction for
positive
eigenvalue:
\begin{equation}
*d*d\hat{\Psi}_{nm}^{(1)} = \omega_{nm}^{2} \, \hat{\Psi}_{nm}^{(1)} \; ,
\hspace{0.5in}
\omega_{nm}^{2} > 0 \; . \label{5.2}
\end{equation}
The subscripts will acquire a meaning similar to that in Section 4, as we shall
see below.
Equations (\ref{5.1}) and (\ref{5.2}) lead to
\begin{equation}
*d*d \hat{F}_{nm}^{(0)} = \omega_{nm}^{2} \, \hat{F}_{nm}^{(0)} \; ,
\label{5.3}
\end{equation}
\begin{equation}
\hat{\Psi}_{nm}^{(1)} = \frac{1}{\omega_{nm}^{2}} \, *d\hat{F}_{nm}^{(0)} \; ,
\label{5.4}
\end{equation}
while (\ref{2.16}) gives
\begin{equation}
\hat{F}_{nm}^{(0)}|_{\partial D} = - \lambda \,
\hat{\Psi}_{nm,\theta}^{(1)}|_{\partial D}
   \label{5.5}
\end{equation}
where
\begin{equation}
\hat{\Psi}_{nm}^{(1)} = \hat{\Psi}_{nm,r}^{(1)} \, dr +
\hat{\Psi}_{nm,\theta}^{(1)} \, rd\theta \; .      \label{5.6}
\end{equation}
Equations (\ref{5.3}-\ref{5.5}) are the analogues of (\ref{4.16}-\ref{4.17}).

As solutions of (\ref{5.3}), we can choose
\begin{equation}
\hat{F}_{nm}^{(0)} = e^{in\theta} \, J_{n}(\omega_{nm}r) \; ,  \label{5.7}
\end{equation}
where, as in (\ref{4.23}), we assume that
\begin{equation}
\omega_{nm} > 0 \; .    \label{5.8}
\end{equation}
The actual eigenvalues are determined by (\ref{5.4}), (\ref{5.5}) and
(\ref{5.6}) which give the
equation
\begin{equation}
e^{in\theta} J_{n}(\omega_{nm} R) =
\{ -\, \frac{\lambda}{\omega_{nm}^{2}} \, \epsilon_{ij} \frac{1}{R}
\frac{\partial x_{i}}{\partial \theta}
 \, \partial_{j} \, [e^{in\theta} J_{n}(\omega_{nm}r)] \} |_{r=R}
 \label{5.9}
\end{equation}
or
\begin{equation}
\lambda =  \frac{\omega_{nm} J_{nm}(\omega_{nm}R)}{J_{n}'(\omega_{nm}R)}
\label{5.10}
\end{equation}
where
\begin{equation}
J_{n}'(\omega_{nm}r) = \frac{d}{d(\omega_{nm}r)} J_{n}(\omega_{nm}r) \; .
\label{5.11}
\end{equation}

Figures 1 to 3 give plots of
\begin{equation}
G_{n}(\omega R) =  \frac{\omega R J_{n}(\omega R)}{J_{n}'(\omega R)}
\label{5.12} \end{equation}
versus $\omega R$ for $n=$ 0, 1 and 10 respectively. We can identify lines of
constant $\lambda R$ in
these figures with lines parallel to the abscissas, the $\omega R$-coordinates
of their
intersections with the $G_{n}(\omega R)$ versus $\omega R$ curves giving
$\omega_{nm}$. The latter are ordered  as in (\ref{4.24}).
The intersections of the graphs of the functions with the abscissas give the
roots of
$J_{n}(\omega R)$ while the intersections of the vertical lines with the
abscissas
give the roots of $J'_{n}(\omega R)$.

\begin{figure}[p]
\begin{center}\mbox{\psannotate{\psbox{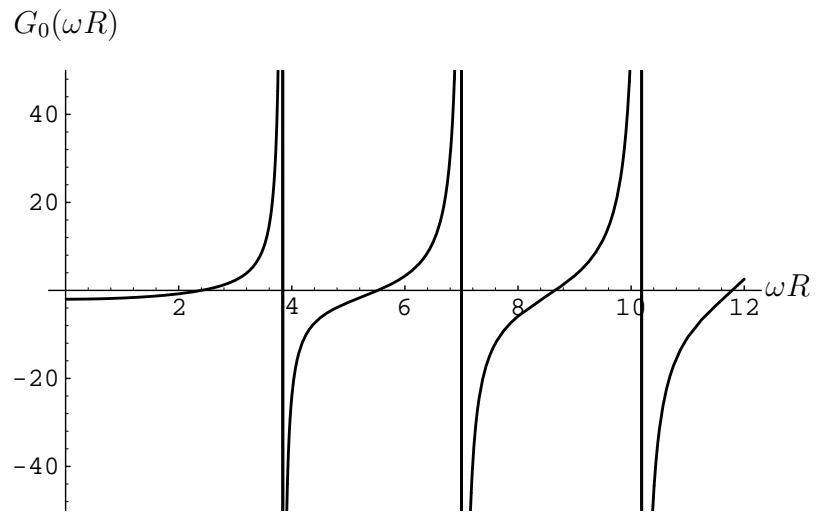}}
{\at(0\pscm;6.5\pscm)
{$G_{0}(\omega R)$}\at(10\pscm;3\pscm){$\omega R$}}}\end{center}
\caption{This figure gives the plot of $G_{0}(\omega R)$ vs $\omega R$.}
\end{figure}
\begin{figure}[p]
\begin{center}\mbox{\psannotate{\psbox{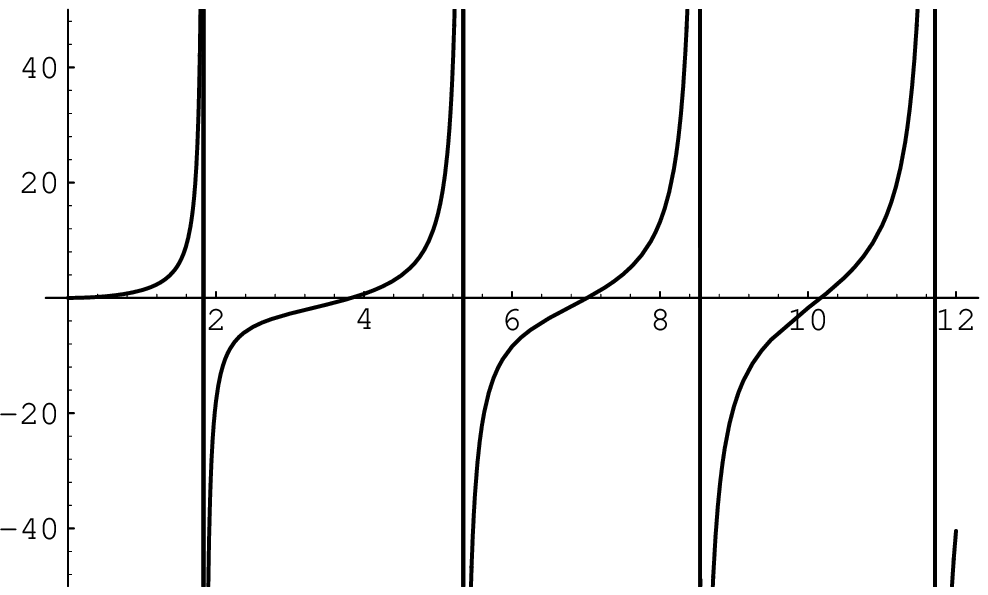}}
{\at(0\pscm;6.5\pscm)
{$G_{1}(\omega R)$}\at(10\pscm;3\pscm){$\omega R$}}}\end{center}
\caption{This figure gives the plot of $G_{1}(\omega R)$ vs $\omega R$.}
\end{figure}
\begin{figure}[hbt]
\begin{center}\mbox{\psannotate{\psbox{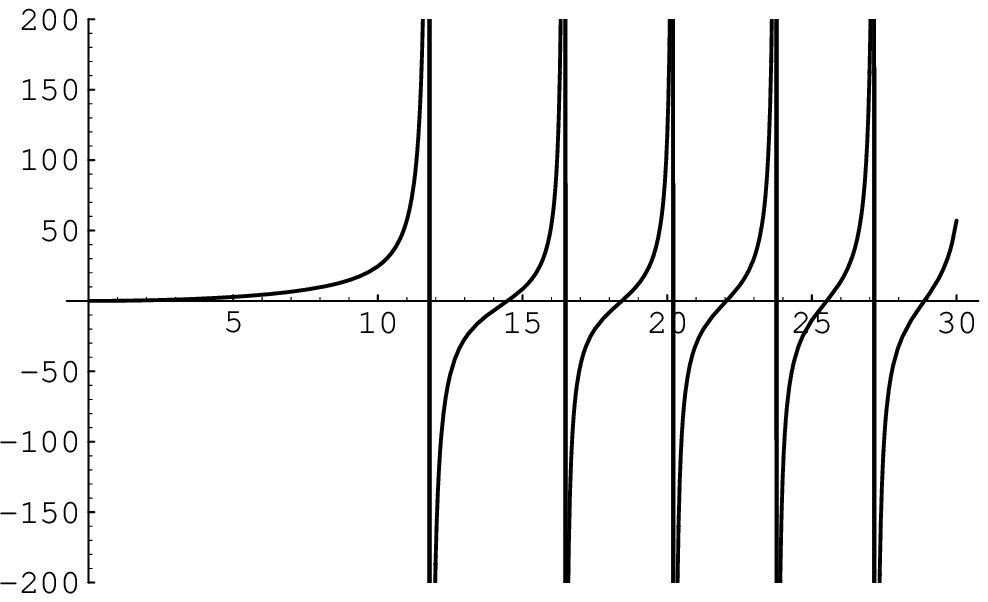}}
{\at(0\pscm;6.5\pscm)
{$G_{10}(\omega R)$}\at(10\pscm;3\pscm){$\omega R$}}}\end{center}
\caption{This figure gives the plot of $G_{10}(\omega R)$ vs $\omega R$.}
\end{figure}

The origins in Figures 2 and 3 (and in general the origin in the graph of the
function $G_{n}(\omega R)$ versus $\omega R$ for $n\neq 0$) merit a few special
remarks. They
correspond to the null
modes of $*d*d$ for $\lambda = 0$ and hence must be associated with the
harmonic
modes $h_{n}^{(1)}, \; \bar{h}_{n}^{(1)}$. The manner in which the
eigenfunctions
acquire this limiting value as $\lambda \rightarrow 0$ is as follows. Let us
first
note that $\hat{\Psi}_{nm}^{(1)}$ is not normalized and introduce the
normalized
eigenfunctions
\begin{equation}
\tilde{\Psi}_{nm}^{(1)} = \tilde{N}_{nm} \, *d\hat{F}_{nm}^{(0)} \; ,
\label{5.13}
\end{equation}
the constants $\tilde{N}_{nm}$ being fixed by
\begin{equation}
(\tilde{\Psi}_{nm}^{(1)},\tilde{\Psi}_{nm}^{(1)}) = 1 \; , \hspace{0.5in}
\tilde{N}_{nm}
> 0 \; . \label{5.14}
\end{equation}
{[ The reason for the tilde here is to distinguish these eigenstates from
those appearing in (\ref{4.29}).]}

Now $\omega_{n1}$ goes to zero as $\lambda$ approaches zero and
$J_{n}(\omega_{n1}r)$ behaves like $[\omega_{n1}r]^{n}$ in that limit. But
since
$\tilde{\Psi}_{n1}^{(1)}$, being normalized to 1, must have a finite value in
that limit, it must be so that $\tilde{N}_{n1} \, \omega_{n1}^{n}$ approaches a
nonzero constant as $\lambda \rightarrow 0$. Hence
\begin{equation}
\tilde{\Psi}_{\pm n,1}^{(1)} \rightarrow h_{n}^{(1)} \, , \, \bar{h}_{n}^{(1)}
\; \hspace{0.5in} \mbox{as} \; \; \lambda \rightarrow 0 \; .  \label{5.15}
\end{equation}
\par

{\bf 5.2 The Hamiltonian and its Eigenstates}\\
\par

The nature of the response of our eigenfunctions of $*d*d$ to the Hodge star
operator had a basic role in the diagonalization of $H$ for $\lambda = 0$. In
that case, the $*$ only permuted these eigenfunctions. Because of this
fortunate
circumstance, the Gauss law could be expressed in a very simple way in terms of
creation-annihilation operators appropriate for the diagonalization of the
Hamiltonian. [This property of $*$ was the reason for the simplicity of the
edge
observables in terms of these operators as well.]

Unfortunately, we can not find eigenfunctions of $*d*d$ with simple
transformation laws under Hodge star when $\lambda$ deviates from zero. As a
consequence, we are unable to handle the commutation relations, the Hamiltonian
$H$ and the Gauss law ${\cal G} (\Lambda^{(0)})$ all at once in a satisfactory
way.
The mode analysis which simplifies the first item of this list is one where
$\Pi$ and $A$ are expanded in a series of our eigenfunctions. But then both $H$
and ${\cal G} (\Lambda^{(0)})$ contain $*A$ in addition to $\Pi$ , and
therefore
get complicated in this mode expansion. Our work on the diagonalization of $H$
and on the Gauss law for nonzero $k$ and $\lambda$ is therefore quite
incomplete.

We can however appreciate certain qualitative aspects of the system already
mentioned in
the Introduction despite the incompleteness of our work. The edge modes
(\ref{4.49})
and their states cease to exist for $\lambda > 0$ because $h_{n}^{(1)}$ and
$\bar{h}_{n}^{(1)}$ which went into their definition in (\ref{4.49}) are not in
the
domain $\cal D_{\lambda}$ for $\lambda \neq 0$. The generators of the group
$\tilde{L}U(1)$ also being these modes for $\lambda=0$, their construction can
not be generalized compatibly with the domain of $*d*d$ for $\lambda>0$.
Although $q_{n}$, $q_{n}^{\dagger}$ still formally generate $\tilde{L}U(1)$,
the
action of the latter probably fails to preserve the domain of the Hamiltonian
because of the aforementioned property of $h_{n}^{(1)}$ and
$\bar{h}_{n}^{(1)}$.
Finally, we already identified the eigenfunctions of $*d*d$ for $\lambda>0$
which become the harmonic modes $h_{n}^{(1)}$, $\bar{h}_{n}^{(1)}$ as $\lambda
\rightarrow 0$. These modes are neither the null nor the degenerate modes of
$*d*d$ for $\lambda >0$. The deformations $H_{n}(\lambda)$ for $\lambda$
positive of the harmonic modes $H_{n}(0)$ can
not therefore be degenerate in energy.

We defer writing the mode expansion of
$\Pi$ and $A$ in  the basis $\{*\Psi_{nm}^{(1)},\tilde{\Psi}_{nm}^{(1)}\}$,
and the commutators of the modes, to
Section 6.

\sxn{The Maxwell Theory on a Disk}

This theory can be explicitly quantized for any $\lambda$ using the basis
of Section 4 or 5. We will consider the cases $\lambda=0$ and
$\lambda>0$ separately.\\
\par

{\bf 6.1. The Case} $\bf \lambda \bf = \bf 0 $ \\
\par

As $k$ is zero, $\Pi$ and $\frac{1}{e^{2}}E$ are the same and can be expanded
as
\begin{equation}
\Pi =\frac{1}{e^{2}}E = \pi_{nm} \Psi_{nm}^{(1)} + \pi_{nm}^{(*)}
*\Psi_{nm}^{(1)} -
q_{n} h_{n}^{(1)} - q_{n}^{\dagger} \bar{h}_{n}^{(1)}    \label{6.1}
\end{equation}
where we have used (\ref{4.49}). [The ranges of summation for $m$ and $n$ in
(\ref{6.1}) are as in (\ref{4.33}).] It is important to note that $q_{n}$ and
$q_{n}^{\dagger}$ here commute:
\begin{equation}
[q_{n},q_{m}] = [q_{n}^{\dagger},q_{m}^{\dagger}] = [q_{n},q_{m}^{\dagger}]
= 0 \; . \label{6.2}
\end{equation}
The expansion of $A$ follows (\ref{4.33}):
\begin{equation}
A = a_{nm} \Psi_{nm}^{(1)} + a_{nm}^{(*)} *\Psi_{nm}^{(1)} + \alpha_{n}
h_{n}^{(1)} + \alpha_{n}^{\dagger} \bar{h}_{n}^{(1)} \; . \label{6.3}
\end{equation}

The nonzero commutators involving operators in (\ref{6.1}) and (\ref{6.3}) are
\begin{equation}
[a_{nm},\pi_{n'm'}] = [a_{nm}^{(*)},\pi_{n'm'}^{(*)}] = i \delta_{n+n',0}
\delta_{mm'} \; ,\nonumber
\end{equation}
\begin{equation}
[\alpha_{n},q_{m}^{\dagger}] =  [\alpha_{n}^{\dagger},q_{m}] = - i \delta_{nm}
\; .      \label{6.4}
\end{equation}
The Gauss law constraint on any physical state $|\cdot \rangle$ follows from
(\ref{6.1}):
\begin{equation}
{\cal G}(*\Psi_{nm}^{(1)}) |\cdot \rangle = 0 \;\; \mbox{or} \; \;
\pi_{nm}^{(*)} |\cdot \rangle = 0 \; .
 \label{6.5}
\end{equation}
The operators $a_{-nm}^{(*)}$ conjugate to
$\pi_{nm}^{(*)}$
are not observables as they do not commute with ${\cal G}(*\Psi_{nm}^{(1)})$.
They will not occur in the Hamiltonian below.

These mode expansions in conjunction with (\ref{2.2}) for $k=0$ and the
Gauss law (\ref{6.5})
give the quantum Hamiltonian
\begin{equation}
H = \frac{1}{2e^{2}}(e^{4}\pi _{nm}^{\dagger}\pi _{nm}+\omega
_{nm}^{2}a_{nm}^{\dagger}a_{nm}+2e^{4}q_{n}^{\dagger}q_{n})  \; .  \label{6.6}
\end{equation}
In (\ref{6.6}), we have dropped terms quadratic in $\pi_{nm}^{(*)}$
in view of (\ref{6.5}),
it being understood that $H$ is restricted to act on physical states. We have
also dropped possible additive constants.

The first two groups of terms here become associated with the photon as $R
\rightarrow \infty$. It is the last set of terms which are especially novel for
the disk. Let us discuss them briefly.

For reasons already stated in Section 4, $q_{n}$ and $q_{n}^{\dagger}$ can be
regarded as edge observables. As they commute, we can see from (\ref{6.6})
that they
are all constants of motion just like the modes $q_{n}$ and $q_{n}^{\dagger}$
in
Section 4. They are in this respect similar to those $q_{n}$ and
$q_{n}^{\dagger}$. But they also differ from the previous $q_{n}$ and
$q_{n}^{\dagger}$ in important ways. Thus for example the group they generate
is
abelian and not $\tilde{L}U(1)$. Also as they appear in $H$, their excitations
now
cost energy. As
\begin{equation}
P_{n}^{(+)}=\frac{q_{n}+q_{n}^{\dagger}}{\sqrt{2}} \; , \hspace{0.3in}
P_{n}^{(-)}=\frac{q_{n}-q_{n}^{\dagger}}{i\sqrt{2}} \; ,\label{6.6b}
\end{equation}
have continuous spectra and behave like infinitely many
translations, it is also clear that $H$ has a contiuous spectrum, its each
point
being infinitely degenerate.

The operators $\alpha_{n}$, $\alpha_{n}^{\dagger}$ are conjugate to
$q_{n}^{\dagger}$, $q_{n}$ by (\ref{6.4}).
Hence
\begin{equation}
Q_{n}^{(+)} = \frac{\alpha_{n}+\alpha_{n}^{\dagger}}{\sqrt{2}} \; ,
\hspace{0.3in}
Q_{n}^{(-)} = \frac{\alpha_{n}-\alpha_{n}^{\dagger}}{i\sqrt{2}}   \label{6.7}
\end{equation}
can be thought of as ``position" operators conjugate to the ``momentum"
operators $P_{n}^{(\pm)}$. [All $n$ here are positive.] We have
\begin{equation}
[Q_{n}^{(\epsilon)},P_{m}^{(\epsilon')}] = i \delta_{nm}
\delta_{\epsilon\epsilon '} \; , \hspace{1cm} \epsilon = \pm \; .  \label{6.8}
\end{equation}
The operators for shifting the eigenvalues of $P_{m}^{(\epsilon)}$ can
therefore
be constructed from suitable exponentials made out of $Q_{n}^{(\epsilon)}$.

Since $Q_{n}^{(\epsilon)}$ [just as $P_{n}^{(\epsilon)}$] commutes with all
${\cal G}(*\Psi_{nm}^{(1)})$, it is
observable.

A noteworthy point may now be made: these $Q_{n}^{(\epsilon)}$ are
\underline{not} observables localized at the edge. Excitations of
$P_{n}^{(\epsilon)}$ created from any state \underline{can not} therefore be
regarded as localized at the edge even though $P_{n}^{(\epsilon)}$ themselves
are localized there. This is to be contrasted with what we found in Section
4.\\
\par

{\bf 6.2. The Case} $\bf \lambda \bf > \bf 0 $ \\
\par

For the Maxwell Lagrangian, in contrast to Section 5, the Hamiltonian and Gauss
law can be expressed using $\Pi=\frac{1}{e^{2}}E$ and $A$ without using $*A$.
For this reason,
all the commutators and the Gauss law can be expressed simply using the modes
of
Section 5. Further the physical states compatible with the Gauss law operator
${\cal G}(*\Psi_{nm}^{(1)})$ are easy to find and the Hamiltonian too is
diagonalized by the basis of Section 5.

For the basis of Section 5.1, the expansions of $A$ and $\Pi=\frac{1}{e^{2}}E$
read
\begin{eqnarray}
A & = & a_{nm} \tilde{\Psi}_{nm}^{(1)} + a_{nm}^{(*)} *\Psi_{nm}^{(1)} \; ,
\label{6.9} \\
\Pi = \frac{1}{e^{2}}E & = & \pi_{nm} \tilde{\Psi}_{nm}^{(1)} + \pi_{nm}^{(*)}
*\Psi_{nm}^{(1)} \; ,       \label{6.10}
\end{eqnarray}
the summation over repeated indices being understood, $n$ being summed over all
integers and $m$ only over positive integers.

The reality of $A$ and $\Pi$ implies the equalities
\begin{equation}
\chi_{nm}^{\dagger} = \chi_{-nm} \hspace{0.5in} \mbox{for} \; \;
\chi=a,a^{(*)},\pi,\pi^{(*)} \label{6.11}
\end{equation}
while the Gauss law is the condition
\begin{equation}
{\cal G}(*\Psi_{nm}^{(1)})|\cdot \rangle = 0 \;\; \mbox{or} \; \;
\pi_{nm}^{(*)} |\cdot \rangle = 0  \label{6.12}
\end{equation}
on physical states $|\cdot \rangle$.

As before, the operator $a_{nm}^{(*)}$ is not an observable in view of
(\ref{6.12}) and
will not occur in the Hamiltonian below.

On dropping terms quadratic in $\pi_{nm}^{(*)}$ using (\ref{6.12}), the
Hamiltonian (restricted to act on physical states) reads
\begin{equation}
H = \frac{1}{2e^{2}}(e^{4}\pi _{nm}^{\dagger}\pi _{nm}+\omega
_{nm}^{2}a_{nm}^{\dagger}a_{nm})             \; .               \label{6.13}
\end{equation}

In contrast to Section 6.1, there are no edge observables for $\lambda>0$. Also
the
spectrum of the Hamiltonian in (\ref{6.13}) is discrete and non-degenerate
for generic
values of $\lambda$, whereas we found a continuous, infinitely degenerate
spectrum for $H$ for $\lambda = 0$.

We conclude the paper repeating the remark that the novel phenomena found here
for the disk seem to have a physical interpretation along the lines outlined in
Section 1. We hope to discuss this interpretation in detail in a paper under
preparation.

\vskip 2cm
\begin{Large}
\begin{center}
{\bf Acknowledgements}
\end{center}
\end{Large}

We thank Giuseppe Bimonte, Dimitra Karabali, Paulo Teotonio-Sobrinho and Ajit
Mohan Srivastava for several discussions. One of us (T.R.G.) will also
like to thank ICTP Trieste for support in the form of associateship and
hospitality at Trieste
during which part of the work was carried out.  We are especially thankful
to Paulo for his help in drawing the figures. The work of A.P.B., L.C. and E.E.
was supported by the Department of Energy, U.S.A., under contract number
DE-FG02-85ER40231.


\begin{thebibliography}{99}

\bibitem{1} J.F. Schonfeld, Nucl. Phys. \underline{B185} (1981) 157; S.Deser,
R.Jackiw and S. Templeton, Phys. Rev. Lett. \underline{48} (1982) 975; Ann.
Phys. \underline{140} (1982) 372.
\bibitem{2} T.J. Allen, M. Bowick and A. Lahiri, Mod. Phys. Lett.
\underline{A6}
(1991) 559; J.A. Minahan and R.C. Warner, Florida Preprint UFIFT-HEP-89-15
(1989).
\bibitem{3} B.I.Halperin, Phys. Rev. \underline{B25} (1982) 2185.
\bibitem{4} A.P.Balachandran and A.M.Srivastava, Minnesota and Syracuse
University preprint TPI-MINN-91-38-T and SU-4228-492 (1991) [to be published by
World Scientific in a volume edited by L.C.Gupta and M.S.Multani].
Talk presented by T.R.Govindarajan on "Edge states in Chern-Simons theory and
quantum Hall systems" at the workshop on "Common problems in Low dimensional
field theories and Condensed matter physics", Madras 5-12 February 1993.
\bibitem{5} A.P.Balachandran, G.Bimonte, K.S.Gupta and A.Stern Int. J. Mod.
Phys. A \underline{7} (1992) 4655, 5855 and references therein;
A.P.Balachandran, "Gauge Symmetries, Topology and Quantization", Syracuse
University preprint SU-4240-506(1992) [Hep-th 9210111] and Proceedings of the
Summer
Course on ``Low Dimensional Quantum Field Theories for Condensed Matter
Physicists", International Centre for Theoretical Physics, Trieste, 24
August to 4 September, 1992 [World Scientific, in press].  Edge states of
the BF system in 3+1 dimensions are discussed in A.P.Balachandran and
P.Teotonio-Sobrinho, Int. J. Mod. Phys. A \underline{8} (1993) 723 and Syracuse
University preprint SU-4240-516(1993) [Hep-th/9304067(1993)] (to be published);
A.P.Balachandran, G.Bimonte and P.Teotonio-Sobrinho, Syracuse University
preprint SU-4240-429 (1992) [Hep-th/9301120(1993)] and Mod. Phys. Letters A (in
press). Analogues of edge states occur also in models based on coset spaces as
shown by T.R.Govindarajan and R.Shankar (unpublished).
\bibitem{6} A.P.Balachandran, L.Chandar, E.Ercolessi and P.Teotonio (in
preparation).
\bibitem{7} P.Goddard and D.Olive, Int. J. Mod. Phys. A1 (1986) 303 and
references therein.
\bibitem{8} A.Pressley and G.Segal, ``Loop Groups'' [Clarendon Press, 1986] and
references therein.
\bibitem{9} J.G.Esteve, Phys.Rev. \underline{D34} (1986) 674.
\bibitem{10} R. Abraham, J.E. Marsden and T. Ratiu, ``Manifolds, Tensor
Analysis and Applications" [Springer-Verlag, 1988].
\bibitem{11} P.A.M.Dirac, ``Lectures on Quantum Mechanics'', Belfer Graduate
School of Science Monographs Series No. 2 [Yeshiva University, New York, 1964];
A.P.Balachandran, G.Marmo, B.S.Skagerstam and A.Stern, ``Classical Topology and
Quantum States'' [World Scientific, 1991] and references therein.
\bibitem{12} B. Simon and M. Reed, `` Methods of Modern Physics" [Academic
Press, 1972] Vol.\ 1.
\bibitem{13} A. Gray and T.M. McRobert, ``A Treatise on Bessel Functions and
their Applications to Physics" [Dover, 1966].

\end{thebibliography}
\end{document}